\documentclass[11pt,aps]{article}
\usepackage[sfdefault]{carlito} 
\usepackage{geometry}
\usepackage{setspace}
\usepackage{titlesec}
\PassOptionsToPackage{hyphens}{url}\usepackage{hyperref}
\usepackage{graphicx}
\usepackage{amsmath}
\usepackage{bm}
\usepackage{wrapfig}
\usepackage{xcolor}
\usepackage{authblk}
\usepackage[export]{adjustbox}
\usepackage{subcaption}
\usepackage{tikz} 
\usetikzlibrary{matrix}
\usepackage[numbers,sort&compress]{natbib}
\usepackage{comment}
\usepackage{upgreek}

\titleformat{\section}{\normalfont\fontsize{11}{13}\bfseries\sffamily}{\thesection}{1em}{}
\titleformat{\subsection}{\normalfont\fontsize{11}{13}\bfseries\sffamily}{\thesubsection}{1em}{}
\titleformat{\title}{\normalfont\fontsize{14}{16}\bfseries\sffamily}{}{0em}{}

\title{Discovering High-Strength Alloys via Physics-Transfer Learning}

\author[1]{Yingjie Zhao}
\author[2]{Hongbo Zhou}
\author[1]{Zian Zhang}
\author[3]{Zhenxing Bo}
\author[3]{Baoan Sun}
\author[2,4]{Minqiang Jiang$^\ast$}
\author[1]{Zhiping Xu$^\ast$}
\affil[1]{Applied Mechanics Laboratory, Department of Engineering Mechanics, Tsinghua University, Beijing 100084, China}
\affil[2]{State Key Laboratory of Nonlinear Mechanics, Institute of Mechanics, Chinese Academy of Sciences, Beijing 100190, China}
\affil[3]{Institute of Physics, Chinese Academy of Sciences, Beijing 100190, China}
\affil[4]{School of Engineering Science, University of Chinese Academy of Sciences, Beijing 101408, China}
\affil[$^\ast$]{Corresponding author(s): Zhiping Xu (xuzp@tsinghua.edu.cn), Minqiang Jiang (mqjiang@imech.ac.cn).}

\date{}

\newlength{\tempdima}
\newcommand{\rowname}[1]
{\rotatebox{90}{\makebox[\tempdima][c]{\textbf{#1}}}}

\newcommand\zfig[1]{{\color{violet}#1}}

\usepackage[T1]{fontenc}
\usepackage{array}
\usepackage{booktabs}
\usepackage{float}

\begin{document}

\maketitle

\begin{abstract}
Predicting the strength of materials requires considering various length and time scales, striking a balance between accuracy and efficiency.
Peierls stress measures material strength by evaluating dislocation resistance to plastic flow, reliant on elastic lattice responses and crystal slip energy landscape.
Computational challenges due to the non-local and non-equilibrium nature of dislocations prohibit Peierls stress evaluation from state-of-the-art material databases.
We propose a data-driven framework that leverages neural networks trained on force field simulations to understand crystal plasticity physics, predicting Peierls stress from material parameters derived via density functional theory computations, which are otherwise computationally intensive for direct dislocation modeling.
This \emph{physics transfer} approach successfully screen the strength of metallic alloys from a limited number of single-point calculations with chemical accuracy.
Guided by these predictions, we fabricate high-strength binary alloys previously unexplored, utilizing high-throughput ion beam deposition techniques.
The framework extends to problems facing the accuracy-performance dilemma in general by harnessing the hierarchy of physics of multiscale models in materials sciences.
\end{abstract}


Scientific development has traditionally followed an observation-assumption-modelling paradigm, from Galileo's beam studies to dislocation theory on the strength of materials~\cite{timoshenko1983history}.
This approach works well in low-dimensional spaces where analytical models can capture system complexity.
However, as we move toward multiscale models, the "curse of complexity" arises, making purely analytical solutions increasingly difficult.
Machine learning (ML) and artificial intelligence (AI) offer a promising data-driven alternative, becoming more powerful as data quantity and quality improve~\cite{wang2023scientific}.
ML's conditional ability to interpolate and extrapolate suggests that it can complement traditional theories where direct solutions are impractical.
Additionally, the feature engineering process in ML parallels dimensional analysis, providing a way to uncover and utilise internal correlations in complex data.
This approach has the potential to advance our understanding of multiscale systems and bridge diverse modelling frameworks.

Atomistic simulations, using first-principles methods and empirical force fields, have revolutionised materials science by enabling high-throughput screening and discovery across a wide range of materials~\cite{jain2013MaterProject,saal2013OQMD,fish2021NatMater}.
However, properties that extend beyond equilibrium structures, such as material strength, are often under-reported due to the high computational cost associated with first-principles methods, which require chemical accuracy (\zfig{Figs. 1a-c})~\cite{mansouri2018machine,jin2023atomic,gupta2022NPJCM,liu2022EML,yang2022CRPS,lee2023ACSAMI}.
To model the mesoscale behaviour of material strength, crystal plasticity (CP) theory provides a bridge between atomic-scale dynamics and continuum stress/strain fields, describing plastic flow and hardening behaviors \zfig{(Fig. 1d, and Supplementary Note 1)}~\cite{roters2011CPFEM}.
Key parameters, such as critical resolved shear stress (CRSS), which governs the activation of slip systems, can be thermally dependent or athermal~\cite{courtney2005MechBehavMater}, and are often derived from experimental measurements~\cite{salem2005ActaMater,gong2015ActaMater}.
Theoretically, the Peierls stress ($\tau_{\rm P}$), defined as the minimum shear stress required to move a dislocation in a perfect crystal, serves as a proxy for CRSS and can be obtained through molecular dynamics (MD) simulations \zfig{(Supplementary Note 1)}~\cite{shimanek2022JOM}.

The strain inhomogeneity caused by the dislocations typically spans $10-20$ nm, making the direct calculation feasible only using empirical force fields~\cite{soleymani2014CMS}.
The Peierls-Nabarro (PN) model offers a simplified yet analytical approach for estimating Peierls stress by assuming a sinusoidal interfacial restoring force and a rigidly shifting dislocation.
The dislocation core structure is determined by minimising elastic energies and lattice misfit~\cite{nabarro1947PNmodel}.
The success of this model suggests that Peierls stress is governed primarily by the elastic responses of the crystal and the energy landscape of interfacial slip systems~\cite{bulatov1997PRL,nabarro1997MSEA,lu2000PRB,rodney2017ActaMater}, although material screening and discovery demand chemical accuracy.
The accuracy-performance trade-off in predicting Peierls stress makes it infeasible to screen high-strength alloys from the expansive materials space.

We propose a physics-transfer (PT) framework to predict the Peierls stress for a wide range of metallic alloys at the first-principles level, leveraging elements from validated theoretical models.
This approach combines the accuracy of first-principles methods with the efficiency of force field simulations.
The maps between the Peierls stress and characteristic materials parameters from empirical or machine learning force field (MLFF) MD simulations are trained using neural networks and then applied by feeding DFT-calculated parameters for chemically accurate predictions.
This method effectively transfers physics through the mapping represented by neural networks, addressing the accuracy-performance trade-off.
We demonstrate its potential through integration with computational materials databases for high-throughput screening, leading to the successful synthesis of high-strength, metastable binary alloys previously unreported (\zfig{Fig. 1a-d}).

\section*{Physics-transfer predictions addressing the accuracy-performance dilemma}
To address the accuracy-performance dilemma \zfig{(Supplementary Note 2)}, the PT framework transfers the underlying physics ($\mathcal{P}$) behind the mapping ($\mathcal{M}$) across models with varying fidelities ($\mathcal{F}$), such as atomistic simulations using empirical force fields and density functional theory (DFT) calculations (\zfig{Figs. 1e and 1f}).
The mapping $\mathcal{M}(\mathcal{P},~\mathbf{x},~\mathcal{F}) \rightarrow \mathcal{O}$, predicts objective properties $\mathcal{O}$ from material parameters $\mathbf{x}$, leveraging low-fidelity (LF) models for training and high-fidelity (HF) models for accuracy.

To predict Peierls stress, we use empirical force fields (e.g., embedded atom methods or EAM) with different force-field parameters and  MLFFs to calculate elastic constants, $\gamma$ surfaces, and Peierls stresses.
Feedforward neural networks (FNNs) and pre-trained ResNet models embed latent features of these characteristics, which are then mapped to Peierls stress obtained from force field simulations \zfig{(Supplementary Note 3)}~\cite{he2016resnet}.
This mapping adheres to the physics of the PN model while employing a data-centric, namely, the `\emph{physics transfer}' approach, and can be extended to address progressive complexity beyond PN and force field models.
Digital libraries are constructed for the fcc system (Cu, Ni, Al, Au, Pd, Pt), the bcc system (Fe, Mo, Ta, W), and the hcp system (Ti, Mg, Zr).
The trained model is applied to DFT-calculated data for materials outside the training set, and the predictions are noted as PT-EAM, PT-MLFFs \zfig{(Supplementary Note 4)}.

To assess the accuracy of the PT predictions, we first calculate the Peierls stress directly by utilizing different methods of calculations, including EAM, MLFFs, DFT, PT-EAM, and PT-MLFFs for small systems (annotated as `S', see \zfig{Methods}) of the fcc system that suffer from strong size effects in predicting the plasticity of bulk materials.
The results indicate that well-trained neural networks effectively learn the physical mapping between the Peierls stress and the characteristic elastic and surface parameters (\zfig{Supplementary Note 5}).
The PT-EAM predictions are quantitatively close ($< 48.91\%$) to those from DFT and MLFF calculations.
In comparison, those obtained from the EAM models show a significant deviation of $221.27\%$ from the DFT predictions (\zfig{Fig. 1g} for Cu $\{111\}\langle 110 \rangle$).
The time cost of statistical inference in the PT approach is within several milliseconds on a laptop, which is significantly lower than that of simulations based on DFT, MLFFs, and EAM (\zfig{Fig. 1h}).
These results obtained for small systems successfully demonstrate advantages in the accuracy and efficiency of the PT approach to predict the Peierls stress.

We then consider large models (`L', see \zfig{Methods}) for $3$ crystalline structures (fcc, bcc, hcp) with their associated specific slip systems (\zfig{Figs. 1i, 1k, and 1m}), where direct DFT calculations are intractable.
MD simulations using MLFFs are performed to validate the accuracy of PT predictions from EAM and MEAM models.
The results show good consistency (with errors $e = 12.55\%$, $48.09\%$, $4.30\%$ for Cu $\{111\}\langle\overline{1}10\rangle$, Fe $\{110\}\langle111\rangle$, Ti $\{10\overline{1}0\}\langle11\overline{2}0\rangle$ in prediction, respectively) and superior performance compared to the EAM results with $e = 33.07\%$, $72.02\%$, $13.89\%$ (\zfig{Figs. 1j, 1l, and 1n}).
By comparing the results obtained for the small and large systems, we also noted that the size effects are more significant for the empirical force fields.
The PT framework thus demonstrates high efficiency compared to DFT and MLFF calculations that can mitigate the size effects, and chemically accurate predictions compared to empirical force fields such as EAM and the analytical PN model \zfig{(Figs. 1j, 1l, 1n, and Supplementary Note 6)}.
\zfig{Table 1} summarises the Peierls stresses for different slip systems calculated by EAM, PT-EAM, PT-MLFFs, and MLFFs.

\section*{Model validation, verification and uncertainty quantification}
Peierls stress prediction exhibits uncertainties in different theoretical methods, making uncertainty quantification (UQ) essential for model evaluation and selection.
\zfig{Fig. 1o} presents error maps for various approaches.
EAM-based predictions for small systems (EAM-S) involve both physical uncertainties (due to potentials) and system uncertainties (due to size effects).
DFT-based small system calculations (DFT-S) eliminate physical uncertainties but retain system uncertainties.
EAM predictions for large systems (EAM-L) reduce system uncertainties, but still contain physical uncertainties.
PT-EAM and MLFFs eliminate both types of uncertainties, with PT-EAM offering superior computational efficiency of training and inference.
\zfig{Fig. 1p} quantifies uncertainty decomposition, showing that EAM-S predictions are dominated by physical and system errors ($99.05\%$), with a small inference error ($0.95\%$), by considering the MLFF results as the ground truth.
For EAM-L, physical and system errors contribute $62.85\%$ and $37.15\%$, respectively.
PT-EAM only has inference uncertainty ($12.55\%$), demonstrating the effectiveness of the PT framework, as supported by machine learning theory (\zfig{Supplementary Note 7})~\cite{abu2012learning,feng2023NMI}.

The accuracy of learned physics is constrained by the fidelity of digital libraries.
While MLFFs offer more accurate physics than EAM or MEAM, their high computational cost and the lack of a complete database for all metal alloys remain challenges~\cite{song2024general}.
Our studies show that PT-MLFFs predictions, using physics learned from MLFF simulations, reduce the error to $e = 1.51\%$ (\zfig{Fig. 1p}).
This few-shot fine-tuning approach significantly improves accuracy over databases constructed with EAM potentials (\zfig{Figs. 1g, 1j, 1l, 1n, and 1p}).

Our PT framework shares conceptual similarities with multi-fidelity ML methods like $\Delta$-learning, low-fidelity-as-a-feature (LFAF), transfer learning, and domain adaption regression which use statistical approaches to combine low- and high-fidelity data for equilibrium properties (see \zfig{Supplementary Note 8} for details)~\cite{ramakrishnan2015DeltaLea,batra2019ACSAMI,smith2019NC}.
Predicting out-of-distribution (OOD) non-equilibrium material properties is still major hurdle with these methods.
The PT framework addresses the challenges by transferring physics attributed to prior insights from predecessors across models of different fidelities using single-point, unit-cell parameters.
For example, in predicting Peierls stresses, key material descriptors such as elastic constants and $\gamma$ surfaces are corrected using DFT or MLFF-trained data \zfig{(Supplementary Note 9)}.
This approach enables accurate, efficient predictions of non-equilibrium properties while balancing accuracy and performance.

\section*{Applications to single elementary metals and alloys}
The high accuracy and efficiency of the PT framework allow a high-throughput screening of single-crystal strength.
For a given material genome, characteristic parameters can be estimated from equilibrium properties in the Materials Project~\cite{jain2013MaterProject} (\zfig{Figs. 2a-c}).
For instance, elastic constants are derived from stress-strain curves (\zfig{Fig. 2a}), while the $\gamma$ surface is fitted from single-point energy calculations (\emph{e.g.}, intrinsic stacking fault energy (SFE) $\gamma_{\rm isf}$, unstable SFE $\gamma_{\rm usf}$, aligned SFE $\gamma_{\rm asf}$, and the energies of their intermediate configurations) and interpolated using a Fourier series (\zfig{Figs. 2b, and Supplementary Note 10})~\cite{su2019JAP}.
The predicted Peierls stresses can be used for strength screening, a long-sought goal in materials science, and extended to mesoscale physics models such as CP (\zfig{Fig. 2c})~\cite{roters2011CPFEM}.

Recent advances in AI-driven materials science have greatly expanded the scope of scientific exploration. 
The graph networks for materials exploration (GNoME) model has increased the inorganic crystal library from $48$k to $2.2$M, including many metastable materials that are difficult to synthesise and assess experimentally using current techniques~\cite{merchant2023Nature}.
The PT model can efficiently screen materials in this vast library at chemically accurate levels, especially for non-equilibrium properties that conventional methods cannot access because of the accuracy-performance dilemma.
For material strength screening via  $\tau_{\rm P}$, we focus on the entire set of $3,471$ fcc (Fm$\overline{3}$m), bcc (Im$\overline{3}$m) or hcp (P$6_3$/mmc) crystals from the GNoME database, supplemented with calculated elastic properties (\zfig{Fig. 2d}).
These crystals have clearly defined slip systems, as opposed to compounds possessing intricate crystal architectures.
A strong linear relationship is found between the SFE and a combined parameter of bulk modulus ($B$), shear modulus ($G$), and lattice constant ($a$) is found, which varies by space group \zfig{(Supplementary Note 11)}~\cite{zhang2023JMRT}.
This SFE is used in our PT approach, bypassing the need to calculate the $\gamma$ surface, which is crucial for accurate $\tau_{\rm P}$ predictions \zfig{(Supplementary Note 11)}.
Subsequently, a comprehensive material strength database is constructed using PT models trained on an ensemble of crystal structures for screening purposes (\zfig{Figs. 2e, 2f, and Supplementary Note 11}), with high-strength metals such as Os, Ru, Tc, and Re showing experimentally verified yield strengths ($\sigma_{\rm Y}$)~\cite{ross2013metallic}, superior to metals in the training set, such as fcc Cu, bcc Fe, hcp Ti (\zfig{Fig. 2g}).

In addition to edge dislocations, screw dislocations also affect material strength, particularly in bcc materials.
Notably, the Peierls stress of bcc materials with compact screw cores can be accurately computed using DFT and MLFFs~\cite{wang2021CMS}.
The PT models trained on individual bcc crystal structures containing screw dislocations also predict Peierls stresses with good consistency ($e = 7.63\%$ for Fe $\{110\}\langle111\rangle$) and superior performance over EAM ($e = 15.01\%$) \zfig{(Supplementary Note 12)}.
In hcp materials, the variation between slip systems is vital for assessing material strength.
The PT models, when trained on distinct hcp crystal structures featuring dislocations within the basal planes, consistently predict Peierls stresses effectively ($e = 41.19\%$ for Ti $\{0001\}\langle11\overline{2}0\rangle$), outperforming EAM prediction with $e = 135.5\%$ \zfig{(Supplementary Note 12)}.

\section*{Prediction and fabrication of high-strength, metastable binary alloys}
Our approach extends to multicomponent alloys or intermetallic compounds incorporating chemical ordering~\cite{ding2018tunable}.
Peierls stress and $\gamma$ surfaces are calculated from pseudorandom structures reproducing the statistical characteristics of atomic arrangements at larger scales~\cite{liu2019MaterDes}.
The physics underlying the strength of alloys goes beyond the PN models.
Specifically, the solid solutions strengthen alloys as the strain field induced by atomic size differences among elements impedes the glide of dislocations.
We integrate strain data from optimized configurations relative to the parent lattice into a neural network, embedding solid solution strengthening physics \zfig{(Supplementary Note 13)}.
This method is demonstrated with binary alloys such as Cu-Ti, Al-Ti, and Al-Mo, known for their excellent corrosion resistance, hardness, and specific strength, making them suitable for aerospace, marine, and electronic uses~\cite{habazaki1997CorrosSci,lee2006Nanotech}.
The PT approach overcomes the limitations of experimental studies on existing materials in exploring metastable solid solutions with high solute content, efficiently screening alloy strengths~\cite{tsuda2004JES}.
Peierls stresses for fcc solid solutions are calculated using PT models and MLFF simulations, showing good consistency (\zfig{Fig. 3a}).
Alloys with higher strength, such as Cu$_{17}$Ti$_{3}$, Cu$_{4}$Ti, Al$_{4}$Ti, Al$_{17}$Mo$_{3}$, Al$_{4}$Mo are identified (\zfig{Fig. 3a}).
The calculation of phase diagrams (CALPHAD) analysis evaluates the stability of these solutions, explaining why high-strength alloys have not been synthesised using traditional methods (\zfig{Fig. 3a, and Supplementary Note 14})~\cite{saunders1997almo}.

PT learning enables the synthesis of high-strength metastable alloys through non-equilibrium alloying and advanced processing techniques.
We perform high-throughput experiments to fabricate solid solute crystals with a wide compositional range via ion beam deposition (IBD)~\cite{zhou2024AFM} (\zfig{Figs. 3b, 3c}, see \zfig{Methods} and \zfig{Supplementary Note 15} for more details).
Metastable alloys with high solute content, such as Cu$_{17}$Ti$_{3}$, Al$_{4}$Ti, are discovered by micro-pillar mechanical tests, achieving a $20$-fold increase in strength over traditional alloys with low solute content.
The plastic flow characteristics measured from micropillar compression tests and the increased dislocation density characterized by inverse fast Fourier transform (IFFT) of transmission electron microscopy (TEM) images after deformation reveal plasticity mediated by dislocation gliding in the metastable crystals, contrasting with the brittle fracture behavior of glass materials (\zfig{Figs. 3d, 3e}).
The metastable crystalline alloys synthesized (Cu$_{17}$Ti$_{3}$ and Al$_{4}$Ti), identified via PT and validated through high-throughput experiments, demonstrate exceptional strength surpassing various materials documented in contemporary research.
This includes Cu- and Al-based alloys, metallic glasses, and intermetallic compounds (\zfig{Fig. 3f})~\cite{zhang2019FED,wang2022MSEA,mao2018MSEA,li2020JAC,sarma2008MSEA,jiang2012JEM,wang2005JNCS,li2018AM}.
Although the PT approach predicts Cu$_{4}$Ti as a high-strength alloy crystal (\zfig{Fig. 3a}), limitations of our IBD technique yield glassy structures instead, which do not display the anticipated high strength (\zfig{Supplementary Note 15}).

\section*{Physical insights from latent space feature analysis}
The physics behind data across different fidelities shapes the neural network structures (\zfig{Fig. 4a}).
Networks trained on low-fidelity empirical force fields and high-fidelity MLFFs show similar parameter distributions, indicating effective transfer of crystal plasticity physics across varying fidelities (\zfig{Supplementary Note 16}).
In predicting Peierls stress for alloys, PT learning leads to neural network weights that are $20.08$\% smaller in magnitude than those derived from statistical learning (\zfig{Fig. 4b}).
Here, `statistical learning' refers to models where the inputs include lattice constants, elastic constants, and the $\gamma$ surface, which do not account for the physics of solid solution strengthening.
Conversely, PT learning integrates an extra input, the average atomic strain, for the essential physics.

Neuron activation rates of PT learning are lower (\zfig{Fig. 4c}) in comparison to statistical learning that does not incorporate the physics.
These characteristics resemble regularization and dropout methods, implying that integrating physics could bolster the generalization of models~\cite{srivastava2014JMLR}.
Understanding the PT framework provides valuable insight for crafting ML algorithms that generalize better.

Differences in physical mechanisms and parameter deviations are reflected in the latent space (\zfig{Fig. 4d}).
Even when model parameters vary, if the underlying physical mechanisms remain constant (such as predicting Peierls stress with MD vs. DFT parameters), the activated neuron positions are unchanged, though their intensities differ (\zfig{Fig. 4e}).
The result clearly approves the assumption behind the PT approach that physical principles can be transferred between models of varying fidelity, and the prediction can be refined by supplying high-fidelity input parameters.
Conversely, when the physical mechanisms are distinct (e.g., predicting Peierls stress for multi-component alloys compared to single-element metals), a greater number of neurons become active, and the differences in activation are more noticeable (\zfig{Fig. 4f}).
These variations aid in evaluating the predictive risks associated with physics-transfer learning.

Examining the relationship between active neurons and data via methods like symbolic regression or dimensional analysis can reveal insights into the fundamental physics (\zfig{Fig. 4g}).
In our study, we detected a notable correlation between the principal eigenvalues of activated neurons and the dislocation width, an essential parameter in the PN model that pertains to crystal plasticity (\zfig{Fig. 4h, and Supplementary Note 16}).
This indicates that the model has inferred the physical principles governing crystal plasticity from the data.

Our study highlights the role of ML-based scientific modeling in addressing data complexity challenges within materials science, exemplified by alloy strength prediction where the nonlocal and nonequlibrium physics is transferred in a data-driven framework.
The PT approach integrates the physics of force field simulations with the chemical accuracy of DFT data using neural network representations, significantly improving both the accuracy and efficiency of predictions.
The demonstrated results pave the way for developing robust metastable alloys by enabling model generalization grounded in physical principles. However, further advancements in scalable material fabrication and processing technologies are required to fully realize the potential of the screened results in engineering.

\clearpage
\newpage

\section*{Methods}

\subsection*{MD simulations}
To construct the digital libraries, a wide spectrum of metals with crystalline structures of fcc (Cu, Ni, Al, Au, Pd, Pt), bcc (Fe, Mo, Ta, W), and hcp (Ti, Mg, Zr) is explored.
The elastic constants, $\gamma$ surfaces, and Peierls stresses are calculated using empirical force fields such as EAM and MEAM with parameters reported from different sources~\cite{becker2013EAMdata,hale2018EAMdata}, as well as the lattice mismatch energy and slip resistance.
The primary slip systems of fcc ($\{111\}\langle\overline{1}10\rangle$) and bcc ($\{110\}\langle111\rangle$), and the prismatic slip systems of hcp ($\{10\overline{1}0\}\langle11\overline{2}0\rangle$) are considered.

In calculations of the $\gamma$ surfaces, a supercell with lattice vectors of $\mathbf{a}~([1\overline{1}0])$, $\mathbf{b}~([11\overline{2}])$, and $\mathbf{c}~([111])$ for the fcc metal is prepared, which contains $64$ atoms and $32$ atomic layers along the $z$-axis.
A vacuum layer of $30$ \AA~along the $z$-axis is added to avoid interactions between the periodic images of lattice mismatch.
The upper $16$ atomic layers are rigidly shifted relative to the lower $16$ layers progressively along the $z$-axis, and independently in the $x$ and $y$ directions, respectively.
Relaxation of the atomic layers along the $z$ direction is allowed after the displacement.
The $\gamma$ surfaces are constructed using a $31\times31$ grid, 
\begin{equation}\label{eq1}
\gamma (x,y) = \frac{E_{\rm m}(x,y)-E_0}{S},
\end{equation}
where $E_{\rm m}(x,y)$ is the energies of the lattice with a mismatch at different displacements, $\mathbf{d} = (x,y)$, $E_0$ is the energy of the crystal in its equilibrium structure, and $S$ is the area of the slip plane.

For the calculations of Peierls stress, a supercell with $\sim 0.8\times 10^{6}$ atoms ($160~\rm{nm} \times2~\rm{nm}\times40~\rm{nm}$) for large models and a supercell with $244$ atoms ($3.48~\rm{nm} \times0.41~\rm{nm}\times1.90~\rm{nm}$) for small models are prepared.
For the fcc metal, lattice mismatch is created between two half-crystals by shifting along the burgers vector by $\mathbf{a}/\sqrt{2}$.
Subsequent structural relaxation then creates an initial edge dislocation~\cite{bulatov2006CSD}.
Molecular statics calculations are used to calculate the Peierls stress identified as the minimum stress at which the motion dislocation is activated~\cite{lim2015IJP}.
A step-wise strain increment of $10^{-5}$ is applied to the supercell.
For bcc and hcp metals, similar procedures are adopted but along different lattice orientations and for different slip systems.
All MD simulations are performed using the large-scale atomic/molecular massively parallel simulator (LAMMPS)~\cite{plimpton1995LAMMPS}.

\subsection*{DFT calculations}
To validate the hypothesis and feasibility of the PT framework, we directly calculate the Peierls stresses in small systems (`S', with 244 atoms) using Cu as an example.
The DFT calculations are carried out using the Vienna \textit{ab initio} simulation package (VASP) using the projector augmented wave (PAW) method and a plane-wave basis~\cite{blochl1994PRB,kresse1999PRB}.
The generalized gradient approximation (GGA) in Perdew-Burke-Ernzerhof (PBE) parametrization is used for the exchange-correlation energy~\cite{perdew1996PBE}.
A supercell for Cu with sizes of $3.48~\rm{nm} \times0.44~\rm{nm}\times1.90~\rm{nm}$ containing an edge dislocation is prepared by structural relaxation using EAM.
A cutoff energy of $500$ eV is chosen for the plane waves and a $1\times5\times1$ ($k_x\times k_y\times k_z$) Monkhorst-Pack $\mathbf{k}$-grid is used to sample the Brillouin zone for Cu~\cite{monkhorst1976PRB}.
The convergence of self-consistent field (SCF) calculations using the plane-wave cutoff and $\mathbf{k}$-grid meshing is assured to be below $1~{\rm meV/atom}$ \zfig{(Supplementary Note 17)}.
Similar to the setup in MD simulations, a step-wise strain increment of $4\times10^{-3}$ is applied.
The Peierls stress is calculated as the minimum stress at which the dislocation is activated to move.

\subsection*{MLFF calculations}
The neuroevolution-potential (NEP) framework is adopted to develop MLFFs for fcc Cu, Al, bcc Fe, and hcp Ti~\cite{fan2021NEP,song2024NEP}.
The local atomic environments are encoded by two-body (radial) and three-body (angular) descriptors.
A feedforward neural network (FNN) with one hidden layer ($30$ neurons) is used to predict atomic energies from these descriptors.
For systems considering dislocation motion and plastic flow, configurations with applied strain and random perturbation of atomic positions, surfaces, and stacking faults are included in the training set, and the energies of these configurations are labeled using DFT calculations.
Instead of using gradient descent-based back-propagation to update the parameters of neural networks, the separable natural evolution strategy algorithm is implemented in the training process to minimize the relatively complex loss functions~\cite{fan2021NEP,song2024NEP}.
The well-trained MLFFs achieve a prediction accuracy of $<5$ meV/atom in the energy of atoms and $<50$ meV/\AA~in the force on atoms (more details of the MLFF development and validation can be found in \zfig{Supplementary Note 18}).
Atomic simulations using the well-trained MLFFs can accurately predict the $\gamma$ surfaces, which is comparable to the DFT calculations, and process higher computational efficiency.

\subsection*{Physics-transfer models}
PT approach improves model fidelity by transferring physics from empirical force fields to neural networks using parameters derived from first-principles force fields.
Models with different fidelities ($\mathcal{F}$) exhibit distinct parameter distributions, $p(\theta|\mathcal{F})$.
Data ($\mathcal{D}$) with different fidelities (e.g., low fidelity from empirical models and high fidelity from DFT) lead to different distributions, limiting model transferability and extrapolation, that is,

\begin{equation}\label{eq1}
p(\theta|\mathcal{D_{\rm LF}}) \neq p(\theta|\mathcal{D_{\rm HF}}).
\end{equation}

However, if low-fidelity models capture the correct underlying physics, as demonstrated by practices in the research community~\cite{buehler2008atomistic}, we propose integrating this physics ($\mathcal{P}$) behind $\mathcal{D}$ into a "physics-transfer" paradigm to facilitate model extrapolation.

If a physical relationship exists between the features ($\mathbf{x}$) and the target properties ($\mathcal{O}$), 

\begin{equation}\label{eq1}
\mathbf{x} \xrightarrow{\mathcal{P}} \mathcal{O},
\end{equation}

\begin{equation}\label{eq1}
\mathbf{x} \cap \mathcal{O} = \mathcal{D}^{'} \subset \mathcal{D},
\end{equation}

the well-trained ML model can learn the underlying physics, resulting in consistent parameter distributions across different fidelities (\zfig{Supplementary Note 16}), that is,

\begin{equation}\label{eq1}
p(\theta|\mathcal{D}_{\rm LF}^{'}) \approx p(\theta|\mathcal{D}_{\rm HF}^{'}),
\end{equation}

This enables the transferability and extrapolation of models, bridging the gaps between empirical and first-principles-derived force fields.

To learn the mapping between elastic constants, $\gamma$ surfaces, and Peierls stresses, we use a convolutional neural network (CNN) to extract features from $\gamma$ surfaces and a FNN to process elastic properties, merging them in the latent feature space.
Another FNN predicts the Peierls stresses.
Given the limited sample size, we employ ResNet for the CNN, fine-tuned on ImageNet~\cite{he2016resnet}.
The FNN for extracting elastic properties contains two layers with $11$ neurons for inputs ($4$ neurons for lattice types and constants, $5$ neurons for independent constants, and $2$ neurons for ISF and USF energes) and $32$ neurons for extracted latent features, respectively.
Following that, the FNN for predicting Peierls stresses has $3$ layers with neuron numbers of 64, 32, and 1, respectively.
The model is trained using stochastic gradient descent (SGD) with learning rates of $10^{-3}$ for dense layers and $10^{-4}$ for fine-tuning layers~\cite{hardt2016ICML}.

Training the model on an ensemble of crystal structures improves accuracy for certain chemistries, such as fcc Cu and bcc Fe, but results in a slight accuracy decrease for hcp Ti compared to models trained on individual crystal structures \zfig{(Supplementary Note 19)}.
This highlights the PT method’s ability to function as a robust fundamental model across different material types.

\subsection*{Material synthesis and characterization}
Traditional alloying creates stable phases per the phase diagram. Metastable alloy crystals need non-equilibrium methods like ion beam deposition (IBD) with high cooling rates.
Cu-Ti alloy films were made using a high-throughput modular target assembly in an IBD system.
The chamber reached ultrahigh vacuum below $8 \times 10^{-5}$ Pa, followed by the introduction of high-purity argon gas ($99.9995\%$) to maintain the ion beam at $2.4 \times 10^{-2}$ Pa.
The process used a constant ion beam current of $60$ mA. Targets included two semi-circular sectors, one of high-purity copper ($99.99\%$) and the other titanium ($99.99\%$), each $100$ mm in diameter, enabling simultaneous sputtering.
A $4$-inch silicon wafer with ($111$) orientation was used as the substrate to collect sputtered atoms and form the alloy.
Held stationary, the substrate allowed a natural compositional gradient due to positional and angular sputtering differences. The process lasted $4$ hours, resulting in a Cu-Ti film approximately $\rm 2.4~\upmu m$ thick.
The high-throughput preparations of the Cu-Ti alloys validate the effectiveness of PT screening (\zfig{Fig. 3f}).
Therefore, for the Al-Ti alloys, we directly prepare the single composition ($\rm Al_{4}Ti$) identified through PT screening (\zfig{Supplementary Note 15}).

Composition analysis was performed at $249$ test points using energy dispersive X-ray spectroscopy (EDS) equipped on scanning electron microscope (SEM).
A Malvern Panalytical Empyrean X-ray diffractometer (XRD) revealed fully amorphous, amorphous-crystalline, and crystalline states for the $\rm Cu_{4}Ti$, $\rm Cu_{5}Ti$, $\rm Cu_{17}Ti_{3}$ alloys, respectively.
Additional structural analysis used a JEM-ARM$200$ aberration-corrected TEM at $200$ kV.
TEM samples were prepared with the FIB technique (Helios 600i, Thermo Fisher) integrated with a Thermo Fisher SEM. 

For mechanical characterization, micropillar compression tests were conducted using a flat indenter on a nanoindenter in displacement-control mode at a strain rate of $5 \times 10^{-4}$ s$^{-1}$.
Micropillars were fabricated using FIB, with consistent geometry: a diameter of $\rm 1~\upmu m$, height of $\rm 2~\upmu m$, and taper angle less than $2^{\circ}$.

\clearpage
\newpage

\section*{Declaration of Competing Interest}

\noindent The authors declare that they have no competing financial interests.

\section*{Credit authorship contribution statement}

\noindent Z.X. conceived and supervised the research.
Y.Z. and Z.Z. developed the theoretical framework and performed the calculations.
H.Z. and Z.B. carried out the experiments under the supervision of B.S. and M.J..
All authors analysed the results and wrote the manuscript.

\section*{Acknowledgements}
 
\noindent This work was supported by the National Natural Science Foundation of China through grants 12425201, 52090032, and 12125206.
The computation was performed on the Explorer 1000 cluster system of the Tsinghua National Laboratory for Information Science and Technology.


\clearpage
\newpage

\bibliography{main_text}

\section*{Figures and Figure Captions}

\begin{figure}[H]
\centering
\includegraphics[width=\linewidth] {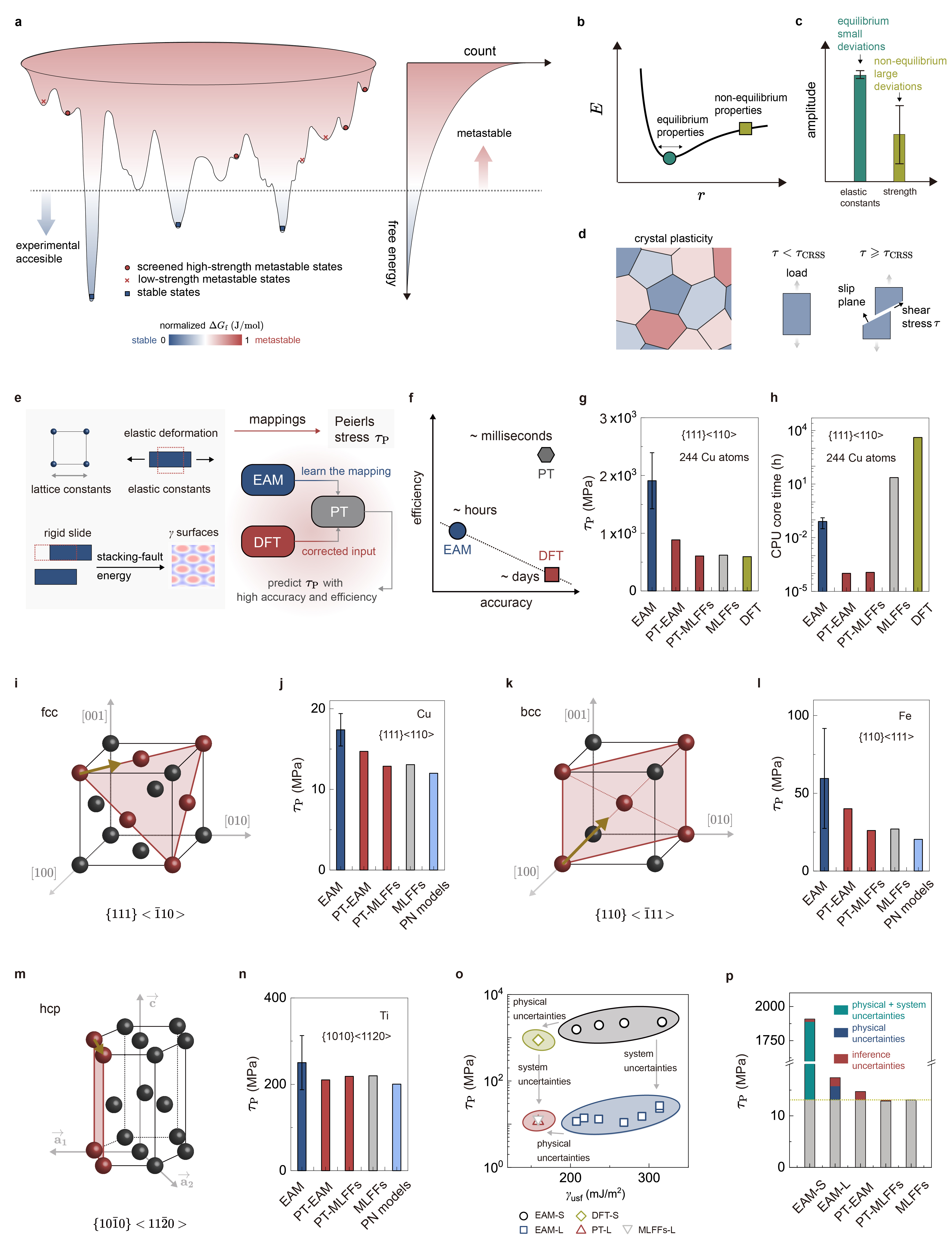} 
\caption{{\bf Predictions of Peierls stress and uncertainties quantification.}
{\bf (a)} The vast metastable materials space and potential high-strength materials.
{\bf (b, c)} Compared to equilibrium properties (e.g., elastic constants), the measurement or calculations of materials’ non-equilibrium properties (e.g., material strength) have larger deviations.
{\bf (d)} Crystal plasticity (CP) and key parameters of critical resolved shear stress (CRSS).
{\bf (e, f)} PT framework transfers the physics from low-fidelity force field models to chemically accurate first-principles methods, effectively addressing the trade-off between accuracy and computational expense.}
\label {fig1}
\end{figure}

\begin{figure}[H]\ContinuedFloat
\caption*{{\bf (g, h)} PT framework predicts Peierls stress with high accuracy and efficiency.
The PT predictions are closely aligned with the outcomes of DFT and MLFF calculations, with a difference below $48.91\%$, while the results obtained using EAM models deviate substantially from the DFT predictions, with a discrepancy of $221.27\%$ {\bf (g)}.
The PT approach also reduces the computational time notably by statistical inference, in comparison with atomistic simulations using DFT, MLFFs, or EAM {\bf (h)}.
{\bf (i-n)} PT predictions for different slip systems ({\bf (i)}: fcc, {\bf (k)}: bcc, {\bf (m)}: hcp).
The PT predictions show good consistency compared to MLFF simulation results (with errors $e = 12.55\%$, $48.09\%$, $4.30\%$ for Cu $\{111\}\langle\overline{1}10\rangle$, Fe $\{110\}\langle111\rangle$, Ti $\{10\overline{1}0\}\langle11\overline{2}0\rangle$ in prediction, respectively), and superior accuracy compared to EAM ($e = 33.07\%$, $72.02\%$, $13.89\%$ ({\bf (j)},{\bf (l)},{\bf (n)}), respectively).
{\bf (o)} Uncertainty quantification shows that the PT predictions eliminate physical and system uncertainties.
`L' denotes the large-supercell system with $\sim 0.8\times 10^{6}$ atoms ($160~\rm{nm} \times2~\rm{nm}\times40~\rm{nm}$).
{\bf (p)} Uncertainty decomposition shows that the inference errors are smaller compared to the physical and system uncertainties.
The standard deviation is reported in the error bars.}
\label{fig1}
\end{figure}

\clearpage
\newpage

\begin{figure}[H]
\centering
\includegraphics[width=\linewidth] {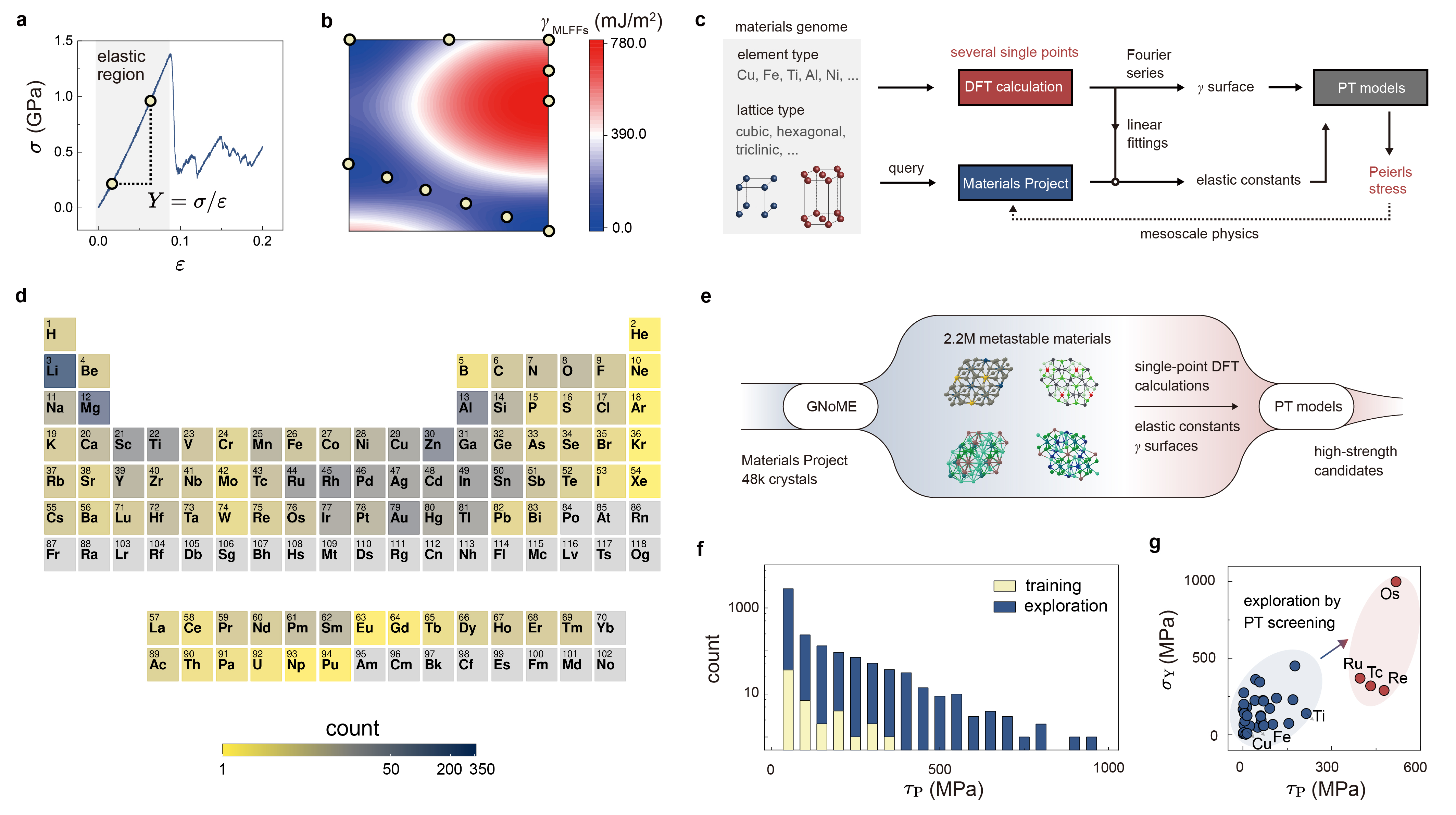} 
\caption{{\bf Material strength screening using PT approach.}
{\bf (a)} Stress-strain ($\sigma$-$\epsilon$) curves under uniaxial tension along the $[100]$ direction of Cu.
The elastic constants ($Y$) are determined as the slope obtained from single-point calculations.
{\bf (b)} The $\gamma$ surface calculated on the $\{111\}$ plane using MLFFs.
The $\gamma$ surface can be fitted by Fourier series on single-point calculations (marked in panel {\bf (b)} as scatter points).
{\bf (c)} The scheme to integrate mesoscale physics into the computational material databases with first-principles accuracy.
{\bf (d)} The material strength database constructed by PT learning, which covers $88$ elements across the periodic table.
{\bf (e)} High-strength material screening from the extensive space of metastable materials in GNoME.
{\bf (f)} The distribution of $\tau_{\rm P}$ in the material strength database.
{\bf (g)} High-strength materials screened using PT learning and the corresponding yield strengths ($\sigma_{\rm Y}$) reported in experiments (extracted from MatWeb~\cite{ross2013metallic}).
}
\label {fig2}
\end{figure}

\clearpage
\newpage

\begin{figure}[H]
\centering
\includegraphics[width=\linewidth] {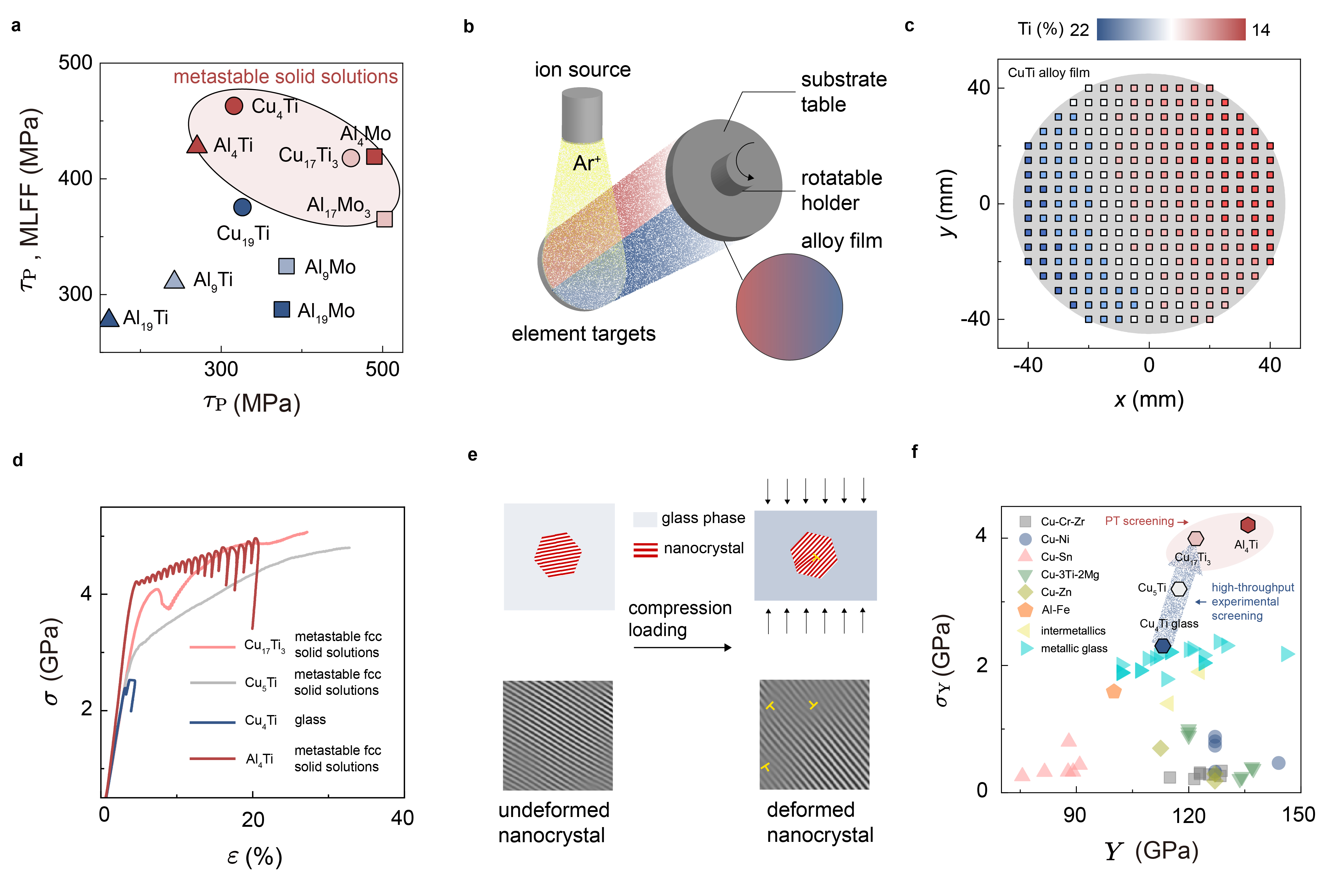} 
\caption{
{\bf PT screening and experimental fabrication of high-strength, metastable alloys.}
{\bf (a)} The metastable binary Cu-Ti, Al-Ti, and Al-Mo alloys with higher strength are screened out.
{\bf (b)} Schematic diagram of the experimental fabrication of metastable alloys using the ion beam deposition (IBD) approach.
{\bf (c)} The Cu-Ti alloy film prepared experimentally has a Ti content ranging from 14 at\% to 22 at\%.
{\bf (d)} The stress-strain curves obtained from microcolumn compression tests for metastable fcc solid solution ($\rm Cu_{17}Ti_{3}$, $\rm Cu_{5}Ti$, $\rm Al_{4}Ti$), and $\rm Cu_{4}Ti$ glass.
{\bf (e)} The heightened dislocation density, as determined by the inverse fast Fourier transform (IFFT) of transmission electron microscopy (TEM) images following deformation, elucidates the deformation mechanism of dislocation glide within the metastable crystal.
{\bf (f)} The screened and fabricated metastable crystal materials ($\rm Cu_{17}Ti_{3}$, $\rm Al_{4}Ti$) using the PT approach and high-throughput experiments demonstrate superior strength in contrast to an extensive array of materials reported in contemporary literature, encompassing Cu- and Al-based alloys, metallic glasses, as well as intermetallic compounds~\cite{zhang2019FED,wang2022MSEA,mao2018MSEA,li2020JAC,sarma2008MSEA,jiang2012JEM,wang2005JNCS,li2018AM}.
For $\rm Cu_{4}Ti$, although the PT approach identifies it as a high-strength alloy crystal, glassy structures are produced in experiments due to constraints associated with IBD.
}
\label {fig3}
\end{figure}

\clearpage
\newpage

\begin{figure}[H]
\centering
\includegraphics[width=\linewidth] {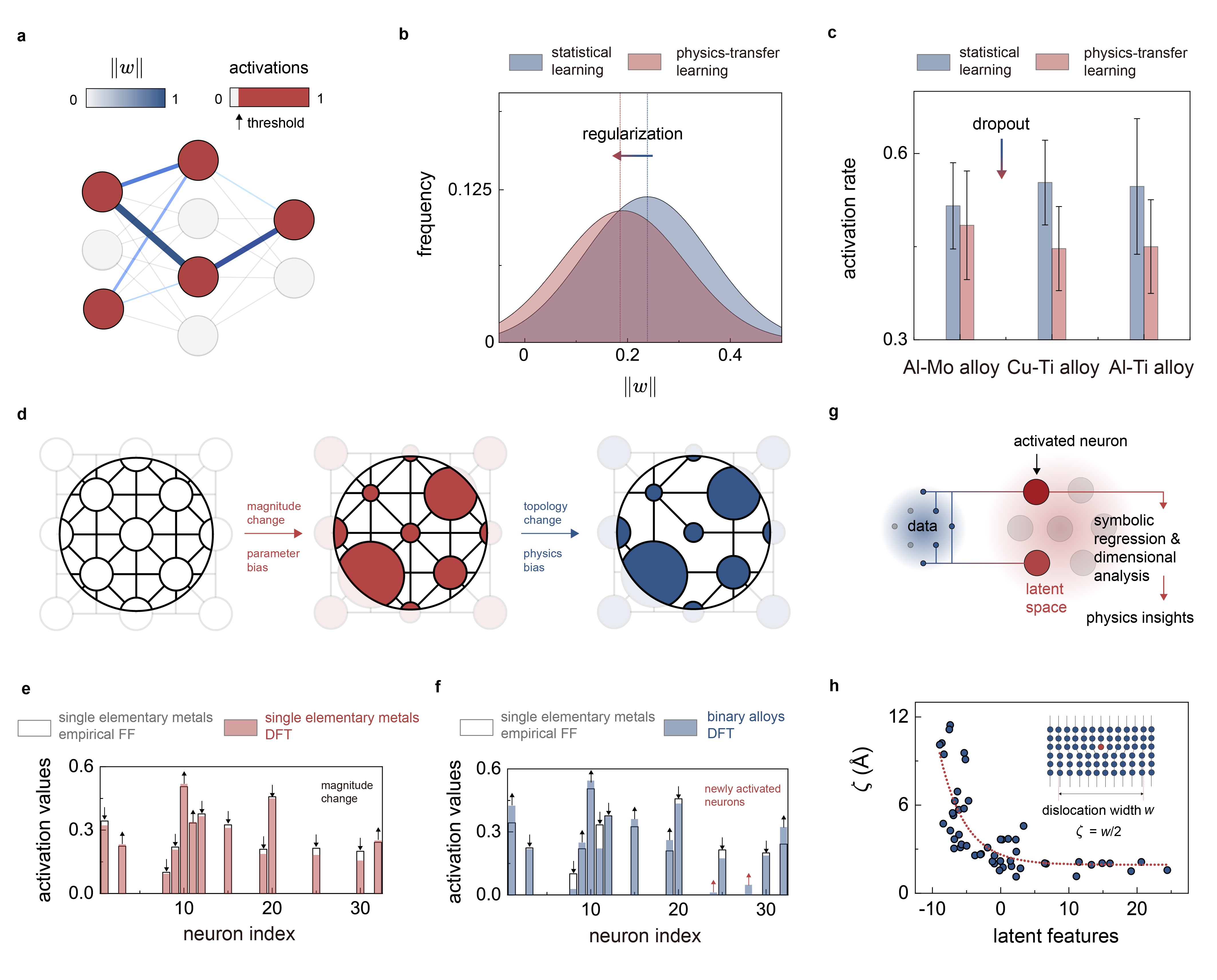} 
\caption{
{\bf Physical insights from latent space feature analysis.}
{\bf (a)} The underlying physics of the data can shape the structures and latent space of the well-trained neural network.
{\bf (b,c)} The neural networks of PT learning possess smaller parameter magnitudes {\bf (b)} and a lower activation rate of neurons {\bf (c)} compared to `statistical learning' with inputs of lattice constants, elastic constants, and the $\gamma$ surface.
In comparison, PT learning also includes the average atomic strain for the physics of solid solution strengthening.
The standard deviation, reported in the error bars, is calculated from $10$ independently trained neural networks.
{\bf (d)} The differences in physics mechanisms and parameter deviations can be reflected in the latent space of neural networks.
{\bf (e)} For the scenarios with the same physical mechanisms but differing parameters, the locations of activated neurons in the latent space are consistent, whereas the intensities of their activation vary.
{\bf (f)} In scenarios where the underlying physical mechanisms differ, additional neurons are activated in the latent space, and the difference in activation magnitudes at the same activation positions becomes more pronounced.
{\bf (g)} Distilling physics by analyzing the relationship between active neurons in the latent space and the data.
{\bf (h)} The strong correlation between the principal eigenvalues of the activated neurons in the latent variable space and the physics of crystal plasticity.
}
\label {fig4}
\end{figure}

\clearpage
\newpage

\begin{table}[H]
  \centering
  \caption{Peierls stress calculated by using EAM, PT-EAM, PT-MLFFs, and MLFFs (the last column is provided for those we developed well-trained MLFFs, see \zfig{Methods} for details).}
  \label{tab:mytable}
  \begin{tabular}{lllll}
  \toprule
  Metal & EAM  & PT-EAM & PT-MLFFs & MLFFs\\
  \midrule
  Cu $\{111\}\langle\overline{1}10\rangle$    & $10.96-28.40$  & $14.71$ & $12.87$  & $13.07$ \\
  Fe $\{110\}\langle111\rangle$ & $38.76-131.09$   & $40.09$ & $26.09$ & $27.01$ \\
  Ti $\{10\overline{1}0\}\langle11\overline{2}0\rangle$  & $170.08-317.36$ & $210.25$ & $218.41$ & $219.70$ \\
  Al $\{111\}\langle\overline{1}10\rangle$    & $6.38-17.71$  & $6.24$ & $9.99$ & $7.41$ \\
  Au $\{111\}\langle\overline{1}10\rangle$    & $4.71-11.61$  & $12.16$ & $12.36$ & $9.43$ \\
  Ni $\{111\}\langle\overline{1}10\rangle$    & $17.18-57.69$  & $18.62$ & $15.01$ & $11.74$ \\
  Pt $\{111\}\langle\overline{1}10\rangle$    & $14.61-35.09$  & $12.86$ & $13.67$ & $11.01$ \\
  Mo $\{110\}\langle111\rangle$ & $50.14-482.28$   & $44.01$  & $98.59$ & $99.48$ \\
  Ta $\{110\}\langle111\rangle$ & $23.25-23.36$   & $35.03$  & $34.34$ & $35.99$ \\
  W $\{110\}\langle111\rangle$ & $45.01-493.57$   & $52.19$  & $87.85$ & $87.42$ \\
  \bottomrule
  \end{tabular}
\end{table}

\end{document}